\documentclass[aps,floatfix,twocolumn,floats,superscriptaddress]{revtex4-1}
\usepackage[T1]{fontenc}
\usepackage{times}
\usepackage{graphicx}
\usepackage{amsmath}
\usepackage{amssymb}
\usepackage{color}
\usepackage{url}
\usepackage{braket}
\usepackage{textcomp}

\usepackage{ulem}

\newcommand{\ham}{\hat{\mathcal{H}}}
\newcommand{\quotes}[1]{``#1''}

\begin{document}


\title{Excited states with selected CI-QMC: chemically accurate excitation energies and geometries}



\author{Monika Dash}
\affiliation{MESA+ Institute for Nanotechnology, University of Twente, P.O. Box 217, 7500 AE Enschede, The Netherlands}
\author{Jonas Feldt}
\affiliation{MESA+ Institute for Nanotechnology, University of Twente, P.O. Box 217, 7500 AE Enschede, The Netherlands}
\author{Saverio Moroni}
\email{moroni@democritos.it}
\affiliation{CNR-IOM DEMOCRITOS, Istituto Officina dei Materiali, and SISSA Scuola Internazionale Superiore di Studi Avanzati, Via Bonomea 265, I-34136 Trieste, Italy}
\author{Anthony Scemama}
\email{scemama@irsamc.ups-tlse.fr}
\affiliation{Laboratoire de Chimie et Physique Quantiques, Universit\'e de Toulouse, CNRS, UPS, France}
\author{Claudia Filippi}
\email{c.filippi@utwente.nl}
\affiliation{MESA+ Institute for Nanotechnology, University of Twente, P.O. Box 217, 7500 AE Enschede, The Netherlands}

\begin{abstract}

We employ quantum Monte Carlo to obtain chemically accurate vertical and adiabatic
excitation energies, and equilibrium excited-state structures for the small, yet
challenging, formaldehyde and thioformaldehyde molecules.  A key ingredient is a
robust protocol to obtain balanced ground- and excited-state Jastrow-Slater wave
functions at a given geometry, and to maintain such a balanced description as we
relax the structure in the excited state. We use determinantal components generated
via a selected configuration interaction scheme which targets the same second-order
perturbation energy correction for all states of interest at different geometries,
and we fully optimize all variational parameters in the resultant Jastrow-Slater
wave functions.  Importantly, the excitation energies as well as the structural
parameters in the ground and excited states are converged with very compact wave
functions comprising few thousand determinants in a minimally augmented double-$\zeta$
basis set.  These results are obtained already at the variational Monte Carlo
level, the more accurate diffusion Monte Carlo method yielding only a small
improvement in the adiabatic excitation energies.  We find that matching Jastrow-Slater
wave functions with similar variances can yield excitations compatible with our
best estimates; however, the variance-matching procedure requires somewhat larger
determinantal expansions to achieve the same accuracy, and it is less
straightforward to adapt during structural optimization in the excited state.

\end{abstract}

\maketitle

\section{Introduction}
\label{sec:intro}

Quantum Monte Carlo (QMC) methods are first-principle approaches that approximately
solve the Schr{\"o}dinger equation in a stochastic manner. The two most commonly
used variants, namely, variational (VMC) and diffusion Monte Carlo (DMC) typically
employ so-called Jastrow-Slater wave functions where a determinant expansion is
multiplied by a Jastrow factor which explicitly depends on the inter-particle
distances and accounts for a significant portion of electronic correlation. Because
of the presence of the Jastrow factor, much shorter expansions are often needed
to describe the Slater component and obtain accurate VMC results, which can then
be further improved with the use of DMC.  Thanks to the favorable scaling with
system size, these methods have been routinely employed to compute the electronic
properties, particularly total energies, of relatively large molecules and
solids~\cite{mitas1996,willow2014,hunt2018,zen2018,chen2018}.

Recent methodological advances have reduced the cost per Monte Carlo step of computing
energy derivatives to the one of the energy itself also for large multi-determinant
wave functions~\cite{filippi2016,assaraf2017}. This has enabled us not only to
simultaneously optimize geometry and wave function with as many as 200,000
determinants but also to explore the dependence of the results on different lengths
and types of Slater expansions, foraying into QMC wave function choices beyond
conventional small-active-space definitions. Capitalizing on these developments,
our recent thorough investigation~\cite{dash2018} of constructing the Slater
component by an automated determinant selection through a selected configuration
interaction (CIPSI) approach~\cite{huron1973,gouyet1976,trinquier1981,
castex1981,pelissier1981,nebot1981,evangelisti1983,cimiraglia1985,cimiraglia1985ab,
cimiraglia1987,illas1991,harrison1991,povill1992} has lead to accurate predictions
of the ground-state energies and structural parameters of butadiene with relatively
short Slater expansions. Relevant determinants could be systematically introduced
into the wave function that would not have been chosen otherwise based on a manual,
intuitive selection.

These developments also open very interesting prospects for the application of QMC
to geometry relaxation in the excited state, where most electronic structure methods
either lack the required accuracy or are computationally quite expensive due to
their scaling with system size. To date, there are very few studies to assess the
ability of QMC to predict excited-state
geometries~\cite{valsson2010,guareschi2013,guareschi2014,zulfikri2016},
while most of the relatively limited literature on excited-state QMC calculations
is primarily concerned with vertical excitation
energies~\cite{schautz2004,schautz2004-2,drummond2005,tapavicza2008,
zimmerman2009,filippi2009,dubecky2010,send2011,filippi2011, valsson2012,valsson2013,
scemama2018,scemama2018excit,pineda2018,blunt2019}.
Importantly, all these studies are characterized by the use of very different wave
functions ranging from the simple ansatz of a CI singles wave function
to complete active space (CAS) expansions, sometimes truncated or only partially
optimized in the presence of the Jastrow factor due to the limitations previously
faced in sampling and optimizing large numbers of determinants.  In this work, we want to
overcome this empiricism in the application of QMC to excited states and a) identify
the most efficient protocol to obtain a balanced and robust description of the
ground and excited states in VMC at a given geometry; b) extend this protocol to
the optimization of the geometry as well as the computation of energy differences
between different potential energy surfaces at distinct geometries; c) demonstrate
the competence of VMC in determining accurate vertical excitation energies, optimal
excited-state structures, and adiabatic excitations.

To this aim, we focus on the low-lying singlet $n\to\pi^{*}$ excitation of
formaldehyde and thioformaldehyde, and compute these excited-state properties
employing compact wave functions containing between a few thousand to about 45,000
determinants obtained through different CIPSI selection schemes.  These molecules are 
small but theoretically challenging: a recent VMC study reported an error as large as 
0.2 eV on the vertical excitation energy of thioformaldehyde~\cite{pineda2018}, and DMC 
calculations without a Jastrow factor required wave functions
with as many as hundred thousand determinants to achieve high accuracy for
formaldehyde~\cite{scemama2018excit}. Here, we obtain VMC vertical excitation energies
within chemical accuracy ($\sim 0.04$ eV) of the extrapolated full CI (FCI) 
estimates for both molecules with expansions as small as a few thousand determinants in 
combination with a minimally augmented double-$\zeta$ basis set.
These excellent results are obtained using CIPSI expansions constructed to yield
a comparable second-order perturbation (PT2) correction in the ground and excited states,
and fully reoptimized in the presence of the Jastrow factor.  Matching the energy
variance of the Jastrow-Slater wave functions of the two states~\cite{robinson2017,pineda2018} to estimate
the excitation energy appears to be a more delicate procedure which, in the case of
formaldehyde, recovers VMC excitation energies within 0.05 eV of our best estimates for
determinantal expansions comprising at least 7000 determinants.

When optimizing the structure, we follow two different selection routes to maintain
a balanced treatment of the wave function while changing the geometry: we construct
the determinantal component targeting a roughly constant value of either the
perturbation correction or the variance of the CIPSI expansions.  We find that
both schemes are viable to obtain robust VMC geometries, also in the more demanding
case of formaldehyde where different correlated methods give a range of variations
in the prediction of the CO bond and the out-of-plane angle in the excited state
as large as 80 m{\AA} and 20$^\circ$, respectively~\cite{budzak2017}. With just a minimal basis,
we obtain optimal VMC structures converged with fewer than a thousand determinants
and in excellent agreement with the coupled cluster estimates, namely, with
deviations smaller than a couple of degrees in the angles and 3 and 10 m{\AA} in
the ground- and excited-state bond lengths, respectively.  Finally, we compute the
difference between the variational minima of the VMC ground- and excited-state
potential energy surfaces to evaluate the adiabatic excitation energies. For both
molecules, irrespective of the determinant selection mode, we obtain VMC and DMC
estimates within better than 0.05 eV of the corresponding coupled cluster values.

The paper is organized as follows. We describe the CIPSI selection scheme employed
to obtain a balanced description of multiple states in Section \ref{sec:methods}
and present the computational details in Section \ref{sec:comput}. The VMC and DMC
vertical excitation energies of formaldehyde and thioformaldehyde are given
in Section \ref{sec:vexc} and the results of the VMC structural relaxation in the
ground and excited states in Section \ref{sec:optgeo}. We conclude in Section
\ref{sec:discuss} by summarizing the most important outcomes of our investigation
and the future prospects of the applicability of our approach.

\section{Methods}
\label{sec:methods}

The wave functions used in the QMC calculations are of the Jastrow-Slater form, namely, the product
of a determinantal component and a positive Jastrow correlation function,
\begin{equation}
\Psi = {\cal J} \sum_k^{N_\textrm{det}} c_{k}D_{k} \,,
\end{equation}
where $N_{\rm det}$ is the total number of determinants and the Jastrow factor $\cal J$ depends here
on the electron-nucleus and electron-electron distances, explicitly ensuring that the
electronic cusp conditions are satisfied.

To construct the determinantal part of the wave function, we employ the CIPSI selected CI algorithm
that iteratively allows us to identify energetically important determinants from the FCI space.  When
one is interested in multiple electronic states, it is important to obtain a description of the CI
subspace which leads to a uniform and balanced treatment of all states of interest.  To ensure a consistent quality
of the wave functions, the selection of the determinants for the multiple states is done in a
single run even if the states belong to different symmetry classes. In practice, starting from an initial
reference subspace $\cal{S}$ typically given by the CI singles wave functions of all states, at every
iteration, we expand the space by selecting among all singly- and doubly-excited determinants those
which contribute the most to a state-average PT2 energy contribution.

If we represent a newly selected determinant with $\ket{\alpha}$, we then
compute its second-order energy contribution using Epstein-Nesbet perturbation theory~\cite{epstein1926,nesbet1955}
as
\begin{equation}
\delta E_{n}^{(2)} = \frac{|\braket{\alpha|\ham|\Psi_n}|^{2}}{\langle \Psi_n|{\hat {\cal H}}|\Psi_n \rangle - \bra{\alpha}\ham\ket{\alpha}}\,,
\end{equation}
and the corresponding CI coefficients as
\begin{equation}
 c_{\alpha, n} = \frac{\langle \Psi_n|\hat{\cal H}|\alpha \rangle}
    {\langle \Psi_n|\hat{\cal H}|\Psi_n \rangle -\langle \alpha |\hat{\cal H}|\alpha \rangle}\,,
\end{equation}
where $n$ denotes a state and $\Psi_n$ is the current CI wave function in the space $\cal{S}$.
As suggested by Angeli {\it et al.}~\cite{angeli1997}, the selection criterion for a determinant
$| \alpha \rangle$ in the external space is based on the energy contribution
\begin{equation}
e_\alpha = \sum_n^{N_\textrm{states}} w_n \delta E_n^{(2)} \,,
\end{equation}
where $w_n$ is the inverse of the maximum $c_{k,n}^2$ with the index $k$ running over the determinants
in the current subspace $\cal{S}$. The determinant $\ket{\alpha}$ is added to $\cal{S}$ if its energy
contribution $e_\alpha$ is higher than a threshold, which is automatically adjusted in such a way that 
the number of determinants in ${\cal{S}}$ is doubled at each iteration.
In the applications presented in this work, the ground state is a closed-shell and the excited 
state an open-shell singlet, both having single-reference character with the dominant configuration 
state function (CSF) comprising a single determinant in the ground state and two determinants 
in the excited state.  Since the expansion is performed in the basis of the determinants 
and not of the CSFs, reweighting the energy contributions by $w_n$ recovers the balance for the different 
states.  As a consequence, this criterion ensures that the multiple states resulting from the expansion 
will have approximately equivalent net PT2 corrections at every CIPSI iteration.

The many-body Hamiltonian, $\ham$, is diagonalized in the subspace $\cal{S}$ at every iteration to
determine the current CI coefficients. One keeps expanding until the wave function has a required
number of determinants or until another chosen criterion is met as for instance a desired total PT2
correction or variance of the CI wave functions as further discussed below in the Result
sections~\cite{dash2018,giner2016, caffarel2016, giner2015,giner2013}.  Additionally, we impose that
the selected wave functions are spin-adapted, namely, eigenfunctions of the ${\hat S}^2$ operator: all
determinants corresponding to the spatial occupation patterns currently present in $\cal S$
are added to the reference space before the Hamiltonian is diagonalized~\cite{applencourt2018,garniron2019}.  
Although wave functions for all the states are obtained through this common selection process, determinants 
of one symmetry character do not contribute, of course,
to a state of a different character.  Therefore, we separate these determinants based on space symmetry
prior to proceeding with the VMC optimization.

In what we name the ``expansion'' scheme, we simply use subsequent wave functions of increasing
length generated according to the CIPSI algorithm we just described.
While the ``expansion'' scheme is more easily transferable to larger systems where one could
experience difficulty in obtaining a large expansion to then truncate, we will also present some tests
with wave functions generated from a ``truncation'' scheme. After having introduced a large
number of determinants in the reference space and obtained the wave functions for both states by
diagonalizing the Hamiltonian, we project the wave functions in a common subspace of determinants as
follows: for each group of determinants, $\mathcal{P}$, of the internal space sharing the same spatial
occupation pattern, we compute the quantity
\begin{equation}
d(\mathcal{P}) = \sum_{k \in \mathcal{P}} \sum_{n}^{N_\text{states}}\, {c_{k n}^2}\,,
\end{equation}
All the individual spatial occupation patterns of the internal space are sorted in decreasing order
of $d(\mathcal{P})$ values, and the truncated determinant space is built by taking the union of the
first patterns of the list.  This guarantees that the truncated wave functions are spin-adapted
and that the final determinants kept after truncation are the most important for all
states of interest.

\section{Computational Details}
\label{sec:comput}

All QMC computations are carried out with the program package
CHAMP~\cite{Champ}.  We employ scalar-relativistic energy-consistent
Hartree-Fock pseudopotentials and the correlation-consistent Gaussian basis sets specifically
constructed for these pseudopotentials~\cite{burkatzki2007,BFD_H2013}.  For the
majority of our calculations, we use a minimally augmented double-$\zeta$ basis
set denoted here as maug-cc-pVDZ and constructed by augmenting the cc-pVDZ basis
with $s$ and $p$ diffuse functions on the heavy atoms.  Basis-set convergence
tests are performed with the fully augmented aug-cc-pVTZ pseudopotential basis.
In both cases, the diffuse functions are taken from the corresponding
all-electron Dunning's correlation-consistent basis sets~\cite{kendall1992}.
The Jastrow factor includes two-body electron-electron and electron-nucleus
correlation terms~\cite{Jastrow}.

We optimize all wave function parameters (Jastrow, orbital, and linear
coefficients) by energy minimization in VMC using the stochastic
reconfiguration (SR) method~\cite{sorella2007} in a conjugate gradient
implementation~\cite{neuscamman2012}. We optimize the ground and excited states
separately since the two states have different symmetries both at the ground-
and the excited-state optimal structures.  We relax the geometry in Z-matrix coordinates
and simply follow the direction of steepest descent, appropriately rescaling the
interatomic forces and using an approximate constant diagonal hessian.
After convergence, we perform additional optimization steps
and average the last 40 structures to estimate the structural parameters
presented below.  To remove occasional spikes in the forces, we use an improved
estimator of the forces obtained by sampling the square of a modified wave
function close to the nodes~\cite{Attaccalite2008}. In the DMC calculations,
we treat the pseudopotentials beyond the locality approximation using the
T-move algorithm~\cite{casula2006a} and employ an imaginary time step of 0.02
a.u. As shown in the Supporting Information (SI), this time step yields DMC
excitation energies converged to better than 0.01 eV for the smallest wave
function in formaldehyde and is therefore appropriate for all wave functions 
of higher quality considered in this work.

The CIPSI calculations are performed with Quantum Package~\cite{scemama2015}
using orbitals obtained from small complete active space self-consistent field
(CASSCF) calculations in the program GAMESS(US)~\cite{schmidt1993,gordon2005}.
As explained above, the CIPSI expansions are constructed to be eigenstates of
${\hat S}^{2}$ and the selected determinants are subsequently mapped into the
basis of configuration state functions, thereby effectively reducing the
number of linear optimization parameters for QMC.

\section{Vertical excitation energies}
\label{sec:vexc}

We begin our investigation by computing the lowest singlet vertical excitations energies of
$n\rightarrow \pi^{*}$ character of formaldehyde (CH$_2$O) and thioformaldehyde (CH$_2$S).
In the ground state, both molecules posses C$_{2v}$ symmetry and the relevant ground (S0) and
excited (S1) states belong to the A$_1$ and the A$_2$ irreducible
representation, respectively.
The QMC vertical excitation energies are computed on the ground-state
structures optimized at the CC3/aug-cc-pVTZ level without the frozen-core
approximation~\cite{scemama2018excit,loos2018}.
Importantly, as we show below, the CC3 geometries are identical to the optimal
ground-state structures obtained at the VMC level.

\subsection{Formaldehyde}
\label{subsec:formaldehyde}

The lowest $n\rightarrow \pi^{*}$ excited state of formaldehyde has been the subject
of a recent DMC investigation~\cite{scemama2018excit}, where Jastrow-free CIPSI
wave functions with as many as 300,000 determinants were employed to recover a
vertical excitation energy within 0.08(3) eV of the best theoretical estimate of
3.97 eV~\cite{loos2018}. In fact, calculations involving about 9000 and 75,000
determinants yielded excitation energies about 0.2 and 0.1 eV higher, respectively,
than the reference value. Given the size of the molecule and the single-reference
character of the excitation, this finding is rather surprising and warrants an
investigation where we generate optimal wave functions in the presence of the
Jastrow factor prior to the DMC step.

\begin{figure}[t]
\includegraphics[width=1.0\columnwidth]{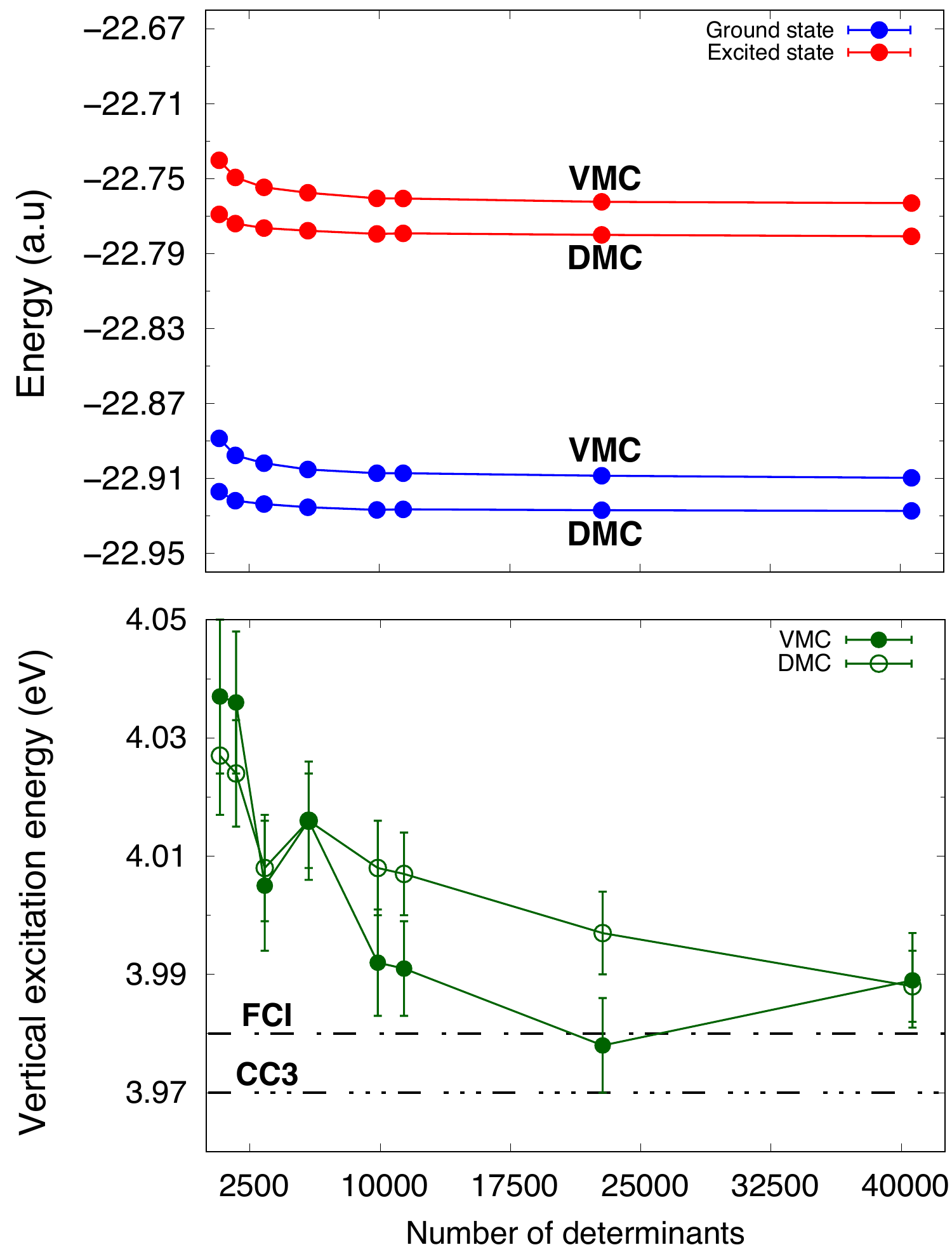}
\caption{Convergence of the VMC and DMC energies of the ground and excited states (top) and of the
excitation energy (bottom) of the $n\to\pi^*$ excitation of formaldehyde with the combined number of
determinants in the CIPSI expansions. The maug-cc-pVDZ basis is used.
We also show the extrapolated FCI and CC3 values obtained with the
aug-cc-pVTZ basis set~\cite{loos2018}. }
\label{fig:form-energy}
\end{figure}

Here, we revisit the VMC and DMC computation of this vertical excitation energy
with compact CIPSI expansions in the Jastrow-Slater wave functions, containing
between 1000 and 40,000 determinants.  These determinant components are the result
of subsequent expansions at the CIPSI level constructed to achieve a balanced
description of the two states of interest at each step by selecting determinants
that yield comparable PT2 contributions for both states as discussed above.  The
determinants corresponding to A$_1$ and A$_2$ symmetry are then isolated and the
Jastrow-Slater wave functions are separately fully optimized by energy minimization
in VMC.  The resulting VMC and DMC total and vertical excitation energies are
listed in Table~\ref{tab:vexc1} and plotted in Fig.~\ref{fig:form-energy}.

\begin{table*}[t]
\caption{VMC and DMC ground- and excited-state energies, and vertical excitation energies (eV) of formaldehyde obtained with
fully optimized Jastrow-CIPSI wave functions, where a series of increasing determinantal expansions are obtained with
the ``expansion'' scheme.
}
\label{tab:vexc1}
\begin{tabular}{lrrrrcccccccc}
\hline
Basis & \multicolumn{2}{c}{No.\ det}  &  \multicolumn{2}{c}{No.\ param}   &     \multicolumn{3}{c}{VMC}                   &&   \multicolumn{3}{c}{DMC}   \\
\cline{6-8}\cline{10-12}
               &        S0         & S1            &   S0    &   S1                    &    E(S0)        &    E(S1)        & $\Delta$E       &&   E(S0)   & E(S1)  & $\Delta$E  \\
\hline
maug-cc-pVDZ &         343          &   436         &   848   &   614           & $-$22.88852(32) & $-$22.74015(33) & 4.037(13)     && $-$22.91702(26) & $-$22.76903(26) & 4.027(10)\\
            &         580          &  1124         &   946   &   961           & $-$22.89762(32) & $-$22.74929(31) & 4.036(12)     && $-$22.92182(25) & $-$22.77393(24) & 4.024(09)  \\
            &         994          &  2360         &  1104   &  1357           & $-$22.90177(30) & $-$22.75459(27) & 4.005(11)     && $-$22.92362(23) & $-$22.77635(23) & 4.008(09) \\
            &        1703          &  4182         &  1375   &  1937           & $-$22.90512(25) & $-$22.75753(25) & 4.016(10)     && $-$22.92530(22) & $-$22.77772(22) & 4.016(08) \\
            &        2747          &  7110         &  1762   &  2805           & $-$22.90717(25) & $-$22.76046(24) & 3.992(09)     && $-$22.92675(22) & $-$22.77947(22) & 4.008(08) \\
            &        3050          &  8320         &  1874   &  3141           & $-$22.90719(20) & $-$22.76051(20) & 3.991(08)     && $-$22.92642(18) & $-$22.77916(17) & 4.007(07) \\
            &        5932          & 16871         &  2915   &  5550           & $-$22.90852(20) & $-$22.76233(20) & 3.978(08)     && $-$22.92687(18) & $-$22.77997(17) & 3.997(07) \\
            &        10854         & 29786         &  4681   &  9177           & $-$22.90961(20) & $-$22.76303(20) & 3.989(08)     && $-$22.92728(14) & $-$22.78072(16) & 3.988(06) \\
aug-cc-pVTZ &         675          &   912         &  3237   &  2256           & $-$22.90151(19) & $-$22.75361(19) & 4.024(07)     && $-$22.92260(16) & $-$22.77499(16) & 4.017(06) \\
            &        1488          &  3214         &  3592   &  3474           & $-$22.91364(17) & $-$22.76638(17) & 4.007(06)     && $-$22.92778(14) & $-$22.78068(13) & 4.015(05) \\
            &        3058          &  7672         &  4210   &  5050           & $-$22.92021(15) & $-$22.77242(15) & 4.021(06)     && $-$22.93112(12) & $-$22.78359(12) & 4.014(05)\\
            &        5849          & 15338         &  5222   &  7401           & $-$22.92050(16) & $-$22.77333(14) & 4.005(06)     && $-$22.93136(11) & $-$22.78360(10) & 4.021(04)\\
            &       12987          & 31710         &  7731   &  12040          & $-$22.92188(15) & $-$22.77466(14) & 4.006(06)     && $-$22.93145(11) & $-$22.78393(13) & 4.014(05)\\
\hline
\end{tabular}
\end{table*}

The VMC vertical excitation energies computed with the maug-cc-pVDZ basis are already within 
0.05~eV of the CC3/aug-cc-pVTZ value with about 1000 and 2400 determinants 
in the ground and the excited state, respectively. Further increasing the size of the wave functions by 
a factor of 10, we observe a rapid convergence to within
$\sim$ 0.02~eV of the CC3 value. The DMC calculations with the VMC optimized wave functions gain about
17-30 mHartrees for the range of expansions studied, so the gain is uniform for both
states.  Consequently, the gap is quite stable also at the DMC level and
consistently compatible within statistical error (to better than 0.02 eV) with
the corresponding VMC values.
A similar behavior is observed when employing the fully-augmented
triple-$\zeta$ basis set with the VMC excitation energy being rather
stable as a function of the expansion size and in agreement
within statistical error with the corresponding DMC value.
From these calculations, we can therefore make
three important observations: a) the weighted CIPSI algorithm used here yields an
automated, balanced determinant selection of the two states; b) in combination with VMC
optimization, we obtain reliable estimates for the vertical excitation
energy with rather small and compact Jastrow-Slater wave functions; c) to
estimate the excitation energy, there is no need to perform a DMC calculation
as the main correlation effects on the energy difference have already been
captured at the VMC level.

The extrapolated FCI estimate of the excitation energy computed with the same
pseudopotential maug-cc-pVDZ basis set is 3.99 eV (see Fig. S1 in the
Supporting Information) in perfect agreement with the all-electron value of
3.99 eV obtained with the corresponding aug-cc-pVDZ basis set. In the
all-electron calculations, the use of the larger aug-cc-pVTZ only reduces the
FCI value to 3.98 eV, that can be further corrected for basis set and
frozen-core errors to yield the best theoretical estimate of 3.97
eV~\cite{loos2018}. Our results of 3.99 and 4.02 eV obtained with the
maug-cc-pVDZ and aug-cc-pVTZ, respectively, are in excellent agreement with the
reference value.

\begin{figure}[t]
\includegraphics[width=1.0\columnwidth]{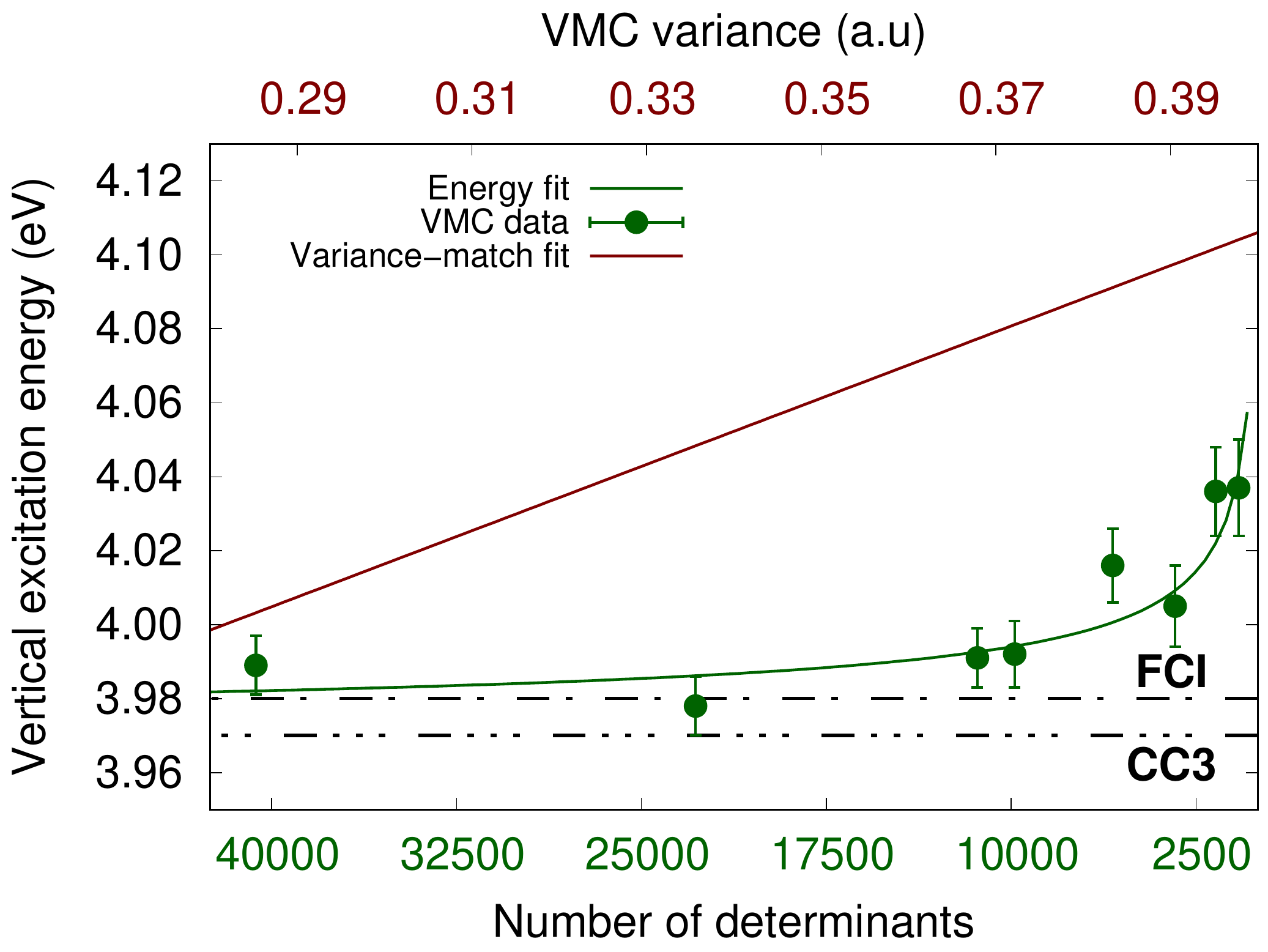}
\caption{
VMC vertical excitation energies (circles) of formaldehyde obtained with the ``expansion'' scheme. The excitation
energy is also estimated as the difference of the fits of the energies against the total number of determinants (bottom x-axis)
and against the variance of the two states (top x-axis). We also show the extrapolated FCI and CC3 values obtained with the
aug-cc-pVTZ basis set~\cite{loos2018}. 
}
\label{fig:fit-form-vexc}
\end{figure}

So far, we have obtained a balanced description of both states by selecting
determinants in the CIPSI expansion that yield comparable net PT2 contribution to both
states at every step.  In Fig.~\ref{fig:fit-form-vexc}, we plot the resulting excitations
energies together with the estimates obtained as difference of the fits of the ground- and excited-state
VMC energies against the number of total determinants as
$\mathrm{E}^\mathrm{fit}_\mathrm{EX}(N_\mathrm{det})-\mathrm{E}^\mathrm{fit}_\mathrm{GS}(N_\mathrm{det})$
(see Section SI.III).
Although we optimize our wave functions separately for the two states in VMC,
we use the total number of determinants as common index since the determinants were obtained from
the same CIPSI selection~\cite{DetTot}.
Furthermore, since we now have at hand multiple wave functions for the ground
and excited states, we can alternatively follow another physically appropriate approach to
compute the excitation energy by matching the variances of the states of interest~\cite{robinson2017,pineda2018}.
To do so, we fit the VMC energies separately for the ground and the excited states against the
corresponding variances
(see Fig. S3) and then subtract the two fits as $\mathrm{E}^\mathrm{fit}_\mathrm{EX}(\sigma^2)-\mathrm{E}^\mathrm{fit}_\mathrm{GS}(\sigma^2)$.  The resulting
variance-matched excitation energies are plotted in Fig.~\ref{fig:fit-form-vexc}, where
smaller variances correspond to larger expansions.

For our fully optimized Jastrow-CIPSI wave functions, both schemes to estimate the
excitation energy yield reasonable values, which become of course more compatible for
the larger expansions.  The excitation energies computed by fitting the energies against the
number of determinants
display a clear and fast convergence, and are closer to the extrapolated FCI estimate over the whole range of expansions
explored here.  For the variance-matching scheme, the fit at high variances deviate from the FCI value
by 0.08-0.12 eV and, only beyond about 7000 determinants and a variance of about 0.31 a.u, the
fit starts giving reasonable estimates within 0.05 eV of our best estimate. Therefore, to make a reliable
variance-matched prediction, one needs to employ larger wave functions
containing several thousand determinants. 
We note that we are not
in a regime \cite{holzmann2015} where a reliable estimate of the exact energy
can be obtained by linear extrapolation to zero variance. 

Finally, while the ``expansion'' scheme represents the most practical route to
generate the determinantal component in a Jastrow-CIPSI wave function for
larger systems, given the small size of formaldehyde, we can also investigate
the effect of using the ``truncation'' scheme where a large CIPSI expansion is
truncated either to yield wave functions with a similar norm for all states
(Table S1, truncation from 40,000 total determinants) or to include the most relevant
patterns (Table S2, truncation from 3 million total determinants) as discussed
in Section \ref{sec:methods}.
We find that both truncated wave functions
yield energetically equivalent estimates of the vertical
excitation energies at the VMC and DMC levels as those obtained with
comparable expansion sizes with the ``expansion'' scheme of
Table~\ref{tab:vexc1}.  We note that the ``truncation'' scheme should be applied to
a relatively large expansion since, during the CIPSI iterations, the largest
coefficients vary much at the beginning but tend to stabilize as the size of
the expansion grows.  If the size of the initial expansion is such that the
determinants kept after truncation have converged coefficients relative to 
the largest coefficient, then truncating the wave function will be independent
of the size of the starting expansion and, therefore, will be equivalent to
truncating the FCI wave function.  In general, fully reoptimizing the trial
wave functions appears to yield accurate and robust estimates of the excitation
energies of formaldehyde for both the ``expansion'' and the ``truncation'' scheme.

\subsection{Thioformaldehyde}

\begin{table*}[htb]
\caption{
VMC and DMC ground- and excited-state energies, and vertical excitation energies (eV) of thioformaldehyde obtained with
fully optimized Jastrow-CIPSI wave functions obtained with the ``expansion'' scheme. The maug-cc-pVDZ basis set is used.
}
\label{tab:vexc2}
\begin{tabular}{ccccccccccc}
\hline
\multicolumn{2}{c}{No.\ det}  &  \multicolumn{2}{c}{No.\ param}   &     \multicolumn{3}{c}{VMC}                         &&   \multicolumn{3}{c}{DMC}   \\
\cline{5-7}\cline{9-11}
 S0         & S1              &   S0            &   S1            &    E(S0)        &    E(S1)        & $\Delta$E       &&   E(S0)   & E(S1)  & $\Delta$E  \\
\hline
  353       &  475            &   854           &    607          & $-$17.04424(26)   & $-$16.96212(25)   & 2.234(10)  && $-$17.06968(25) & $-$16.98671(24) & 2.258(9)\\
  702       & 1488            &   995           &   1044          & $-$17.04998(25)   & $-$16.96814(24)   & 2.227(09)  && $-$17.07246(23) & $-$16.99023(23) & 2.237(9) \\
 1165       & 2702            &  1168           &   1461          & $-$17.05239(24)   & $-$16.97019(24)   & 2.237(09)  && $-$17.07284(22) & $-$16.99086(22) & 2.231(9) \\
 1834       & 4692            &  1419           &   2034          & $-$17.05388(23)   & $-$16.97100(23)   & 2.255(09)  && $-$17.07394(22) & $-$16.99171(22) & 2.237(9)\\
 2500       & 6316            &  1662           &   2514          & $-$17.05494(23)   & $-$16.97260(23)   & 2.241(09)  && $-$17.07434(21) & $-$16.99211(22) & 2.238(8)\\
 3432       & 8338            &  1997           &   3112          & $-$17.05622(23)   & $-$16.97313(23)   & 2.261(09)  && $-$17.07482(22) & $-$16.99219(22) & 2.248(8)\\
 5712       & 14562           &  2810           &   4848          & $-$17.05698(22)   & $-$16.97497(22)   & 2.232(08)  && $-$17.07542(22) & $-$16.99314(26) & 2.239(9)\\
14218       & 30142           &  5650           &   9162          & $-$17.05780(15)   & $-$16.97590(19)   & 2.236(07)  && $-$17.07566(18) & $-$16.99345(26) & 2.237(8)\\
\hline
\end{tabular}
\end{table*}

This small molecule has been the subject of many recent
investigations~\cite{loos2018-2,pineda2018,wiberg2005,de1985} and is a variant of
formaldehyde, wherein the O atom is replaced by a S atom.
Interestingly, the very recent study by Flores {\it et.\ al}~\cite{pineda2018} employed
Jastrow-CIPSI wave functions to compute the
excitation energy of the $n\rightarrow \pi^{*}$ transition but obtained a VMC excitation energy of
2.07(2) eV, that is, about 0.2 eV lower than the reference extrapolated stochastic-heat-bath CI
value of 2.31(1) eV on their geometry. They optimized
all wave function parameters, albeit in variance minimization. In view of our success with
the computation of the vertical excitation energy of formaldehyde at the VMC level, it is
somewhat puzzling that the simple substitution of oxygen with sulfur would worsen so much the performance
of the method. Therefore, we revisit here the same molecule using fully optimized Jastrow-CIPSI
wave functions and both the CIPSI selection procedure of multiple states with
similar PT2 energy correction and the variance-matching scheme.

\begin{figure}[t]
\includegraphics[width=1.0\columnwidth]{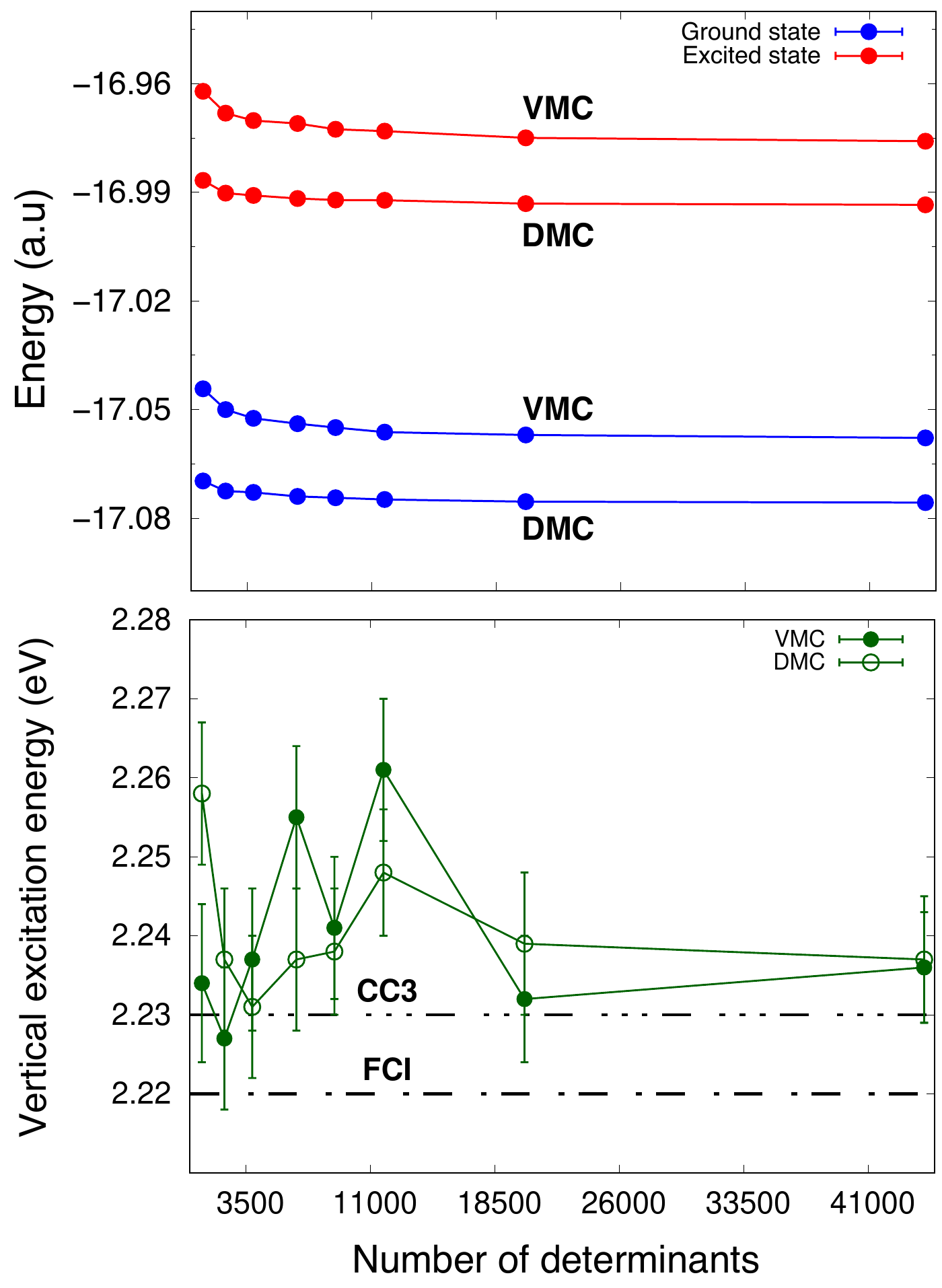}
\caption{Convergence of the VMC and DMC energies with the number of CIPSI determinants of the ground 
and excited states (top) and of the excitation energy (bottom) of the $n\to\pi^*$ excitation of thioformaldehyde.
The maug-cc-pVDZ basis is used.
We also show the extrapolated FCI and CC3 values obtained with the aug-cc-pVTZ basis set~\cite{loos2018}.
}
\label{fig:form-thio-energy}
\end{figure}

We summarize our QMC results obtained from the CIPSI ``expansion'' scheme and the maug-cc-pVDZ basis set
in Table \ref{tab:vexc2} and also present them in Fig.~\ref{fig:form-thio-energy}.
As in the case of formaldehyde, we find that we are able to obtain a stable estimate of the vertical
excitation energy both at the VMC and DMC level with relatively little effort.
The excitation energy is essentially the same within statistical error
at both the VMC and the DMC level when increasing the total number of determinants from
800 to 44,000. The use of larger expansions yields of course a gain in the total energies,
which is however less than 1 and 0.5 mHartree in VMC and DMC, respectively, when doubling
the expansion size between the last two entries of Table \ref{tab:vexc2}.
Computation of the DMC energies on top of the final VMC optimized
wave functions uniformly lowers the energies of both states by about
18-20 mHartrees, thereby unaltering the energy separation between them.
Therefore, the CIPSI selection scheme combined with wave function optimization ensures
a balanced description of the two states and yields
a VMC excitation energy of 2.23 eV in excellent
agreement with the CC3 value of 2.23 eV and the extrapolated FCI estimate of 2.22 eV obtained
with a larger aug-cc-pVTZ basis~\cite{loos2018-2}.

\begin{figure}[!]
\includegraphics[width=1.0\columnwidth]{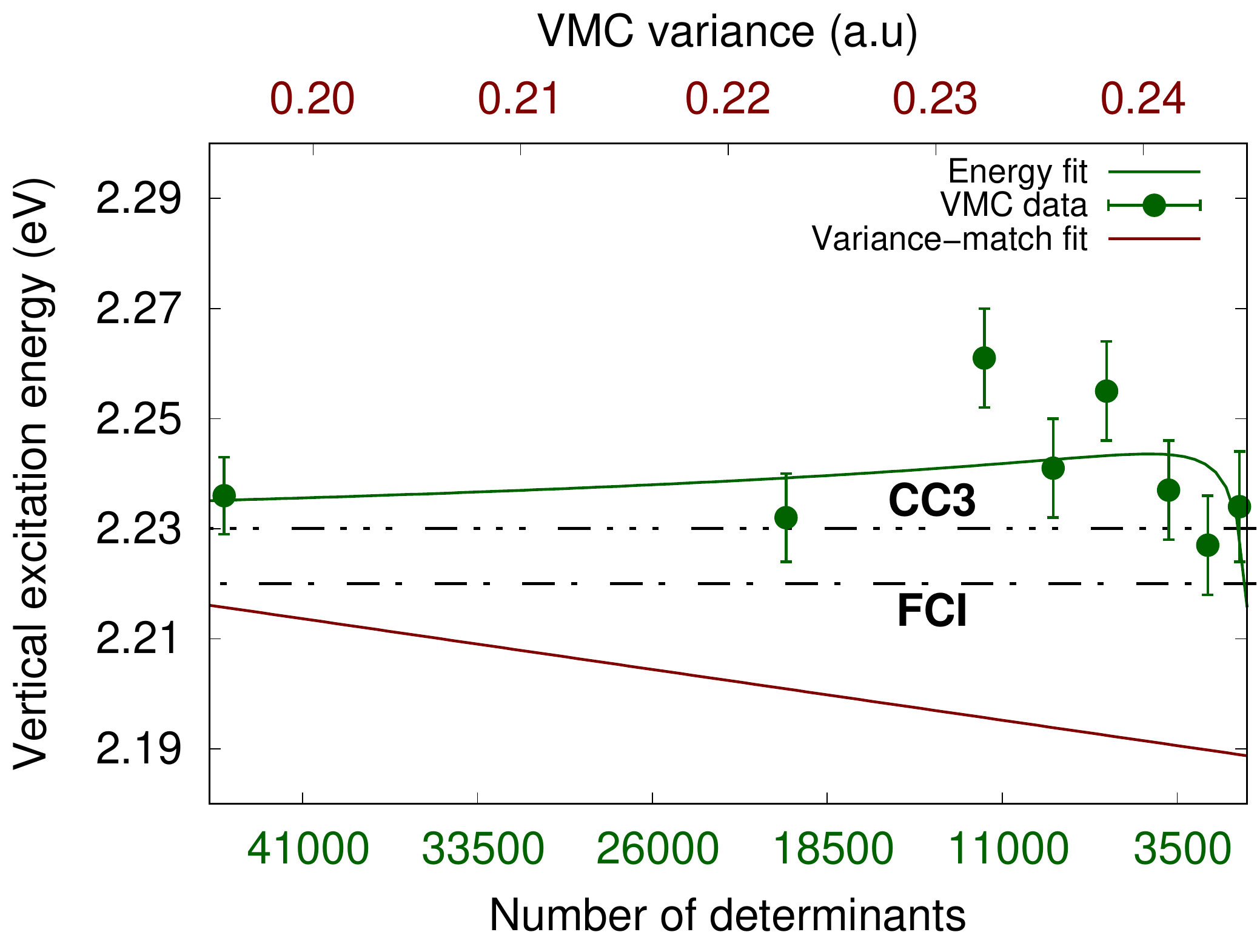}
\caption{
VMC vertical excitation energies (circles) of thioformaldehyde obtained with the ``expansion'' scheme. The excitation
energy is also estimated as the difference of the fits of the energies against the total number of determinants (bottom x-axis) and against the variance of the two states (top, x-axis). We also show the extrapolated FCI and CC3 values obtained with the aug-cc-pVTZ basis set~\cite{loos2018}.
}
\label{fig:thio-variance}
\end{figure}

In Fig.~\ref{fig:thio-variance}, we plot the excitation energies obtained for the different
CIPSI expansions together with the two estimates one obtains by computing the difference of the fits of
the ground- and excited-state energies against the total number of determinants and against the variances of the
two states as done for formaldehyde. We find that both estimates are compatible
well within 0.05 eV over the whole range of expansion sizes/variances. We therefore do not reproduce
the large error of 0.2 eV reported in Ref.~\cite{pineda2018}, which cannot therefore be due to the use
of the variance-matching recipe with reoptimized Jastrow-CIPSI wave functions.

\section{Optimal excited-state structures and adiabatic excitation energies}
\label{sec:optgeo}

Upon relaxation of formaldehyde in the first singlet excited state, the oxygen
atom moves out of the plane. The optimal excited-state structure therefore
possesses C$_{s}$ symmetry with the symmetry plane being perpendicular to the
initial molecular plane and passing through the CO bond. The excited-state optimization
therefore follows the $A^{\prime\prime}$ state as the symmetry of the molecule is lowered from
C$_{2v}$ to C$_{s}$.  In the first singlet excited state, thioformaldehyde remains instead planar
and preserves C$_{2v}$ symmetry so that the excited state maintains A$_2$ character.

\subsection{Formaldehyde}

\begin{table}[b!]
\caption{Optimal VMC ground-state bond lengths (\AA) and bond angles (deg) of formaldehyde obtained using
Jastrow-CIPSI wave functions with the maug-cc-pVDZ basis. The VMC energy (a.u.) is averaged over the last
40 iteration of geometry optimization. We also list the CC3/aug-cc-pVTZ
and the experimental values.
}
\label{tab:form-geo_S0}
\begin{tabular}{cccccc}
\hline
No.\ det.  &   No.\ par.    &   CO       &   CH       &    H-C-H &   $\langle$E$_{\rm VMC}\rangle$  \\
\hline
 580       &    946        & 1.20754(08) & 1.09871(9)  & 117.09(6) &  $-$22.89824(4)  \\ 
 2747      &   1762        & 1.20791(13) & 1.09952(6)  & 116.85(3) &  $-$22.90716(4)  \\ 
 5932      &   2915        & 1.20819(08) & 1.09952(3)  & 117.05(3) &  $-$22.90867(3)  \\ 
\hline
\multicolumn{2}{l}{CC3~\cite{budzak2017}}                 & 1.208       & 1.100        & 116.4   &    \\
\multicolumn{2}{l}{Expt.~\cite{clouthier1983}} & 1.208       & 1.116        & 116.3   &    \\
\hline
\end{tabular}
\end{table}

We first optimize the structure of formaldehyde in the
ground state for some of the Jastrow-Slater CIPSI wave functions that we have
used to compute the vertical excitation energies of Table~\ref{tab:vexc1}.
Similar to our previous findings for butadiene~\cite{dash2018}, we obtain an
accurate geometry in the ground state with relatively short expansions.  As
shown in Table~\ref{tab:form-geo_S0}, the optimal bond lengths and bond angles
computed with the compact aug-cc-pVDZ basis are within 1 m{\AA} and 1$^\circ$ of
the corresponding CC3/aug-cc-pVTZ values already with the smallest wave function
considered. Energetically, we gain about 1 mHartree upon
structure optimization with the smallest wave function of 580 determinants
and the average VMC energies of the two larger cases are identical to the
initial energies on top of the starting CC3 geometry.

To relax the geometry in the excited state, we can simply start from 
the ground-state structure, with the oxygen slightly displaced out of plane, 
and use the same excited-state expansion as in the calculation of the vertical 
excitation.  Even though such a wave function misses all determinants of B$_2$ 
character which will acquire non-zero weight as the molecule moves out of plane,
this quick procedure gives us already a good estimate of the excited-state
geometry with a CO bond of about 1.334 {\AA} and an out-of-plane angle of 31.5$^\circ$
(see Table S4). These structural parameters compare very favorably with the corresponding values
obtained in CC3 and a full valence CASPT2 calculation~\cite{budzak2017}.
We stress that the excited-state geometry of formaldehyde has been found to depend significantly on the level of
theory, with highly-correlated methods spanning differences of as much as 0.08
\AA\ in the CO bond length and 20 degrees in the out-of-plane angle subtended
by this bond on the HCH plane~\cite{budzak2017}.

Starting from such an out-of-plane excited-state geometry, we generate
a series of CIPSI expansions of symmetry A$^{\prime\prime}$ and continue the
structural relaxation in the excited state to further investigate the
dependence of the resultant geometry on the complexity of the wave function.
Since we also want to compute the adiabatic excitation energy, we have to
evaluate differences of the ground/excited-state energies on the ground/excited-state optimal
geometries. To obtain a balanced description, we therefore generate the CIPSI
wave functions at different geometries either targeting a given value
of the PT2 correction or of the CI variance for all four states.  For convenience,
we shall hereafter refer to them as the iso-PT2 and the iso-variance procedure,
respectively. The results of these calculations are shown in Table
\ref{tab:iso-form-geo}.

\begin{table*}[htb!]
\caption{Optimal VMC ground- and excited-state bond lengths (\AA) and bond
angles (deg), and VMC and DMC adiabatic excitation energies (eV) of formaldehyde obtained using Jastrow-CIPSI wave functions with
targeted PT2 correction (iso-PT2). The VMC energy (a.u.) is averaged over the
last 40 iterations of geometry optimization. o.o.p denotes the out of plane
angle of the CO bond.}
\label{tab:iso-form-geo}
\begin{tabular}{lcccccccccc}
\hline
State & $\delta$E$_{\rm PT2}$ &  No.\ det.   &   No.\ par.   &   CO       &   CH        &    H-C-H    &   o.o.p    & $\langle$E$_{\rm VMC}\rangle$  & $\Delta$E$_{\rm adia}^{\rm VMC}$ & $\Delta$E$_{\rm adia}^{\rm DMC}$  \\
\hline
S0 & 0.20     &  515   &   1183   & 1.20663(05) & 1.09851(03)  & 116.84(4) & $-$0.098(25) &  $-$22.89892(5)   \\
   & 0.15     & 1184   &   1496   & 1.20795(08) & 1.09832(05)  & 116.96(6) & \;\; 0.075(15) &  $-$22.90411(6)   \\
   & 0.10     & 2784   &   2208   & 1.20752(19) & 1.09947(05)  & 117.00(5) & \;\; 0.073(54) &  $-$22.90723(5)   \\
   & 0.07     & 4799   &   2946   & 1.20710(14) & 1.09959(10)  & 117.01(8) & $-$0.030(29) &  $-$22.90805(4)   \\
   &\multicolumn{3}{l}{CC3~\cite{budzak2017}}    & 1.208       & 1.100        & 116.4       &  0.000   \\
   &\multicolumn{3}{l}{Expt.~\cite{clouthier1983}}  & 1.208       & 1.116        & 116.3       &  0.000   \\[1.5ex]
S1 & 0.20     & 1088   &   1363   & 1.33948(07) & 1.08468(02)  & 119.83(03)& 32.916(055) &  $-$22.76787(5) & 3.566(2) & 3.612(9) \\
   & 0.15     & 3106   &   2037   & 1.33971(36) & 1.08630(06)  & 119.03(20)& 34.778(051) &  $-$22.77342(5) & 3.556(2) & 3.614(8) \\
   & 0.10     & 8058   &   3520   & 1.33606(21) & 1.08647(12)  & 119.34(17)& 34.027(233) &  $-$22.77626(3) & 3.564(2) & 3.597(8) \\
   & 0.07     & 15278  &   5589   & 1.33597(15) & 1.08645(08)  & 119.56(09)& 33.294(070) &  $-$22.77746(3) & 3.553(1) & 3.592(8) \\
   & \multicolumn{3}{l}{CC3~\cite{budzak2017,loos2018-2}}  & 1.326  & 1.089  & 118.3  &  36.8  &         & 3.602\\
   & \multicolumn{3}{l}{Expt.~\cite{job1969,jensen1982form,godunov1999}} & 1.321--1.323 & 1.092--1.103  & 118.1--121.5  &  20.5--34   \\
\hline
\end{tabular}
\end{table*}

\begin{table*}
\caption{
Same as Table~\ref{tab:iso-form-geo} using Jastrow-CIPSI wave functions with
	targeted CI variance (iso-variance).}
\begin{tabular}{lcccccccccc}
\hline
State & $\sigma_{\rm CI}^2$ &  No.\ det.   &   No.\ par.   &   CO       &   CH        &    H-C-H    &   o.o.p    & $\langle$E$_{\rm VMC}\rangle$  & $\Delta$E$_{\rm adia}^{\rm VMC}$ & $\Delta$E$_{\rm adia}^{\rm DMC}$  \\
\hline
S0  & 0.80 &  515 &   1183        & 1.20663(05) & 1.09851(03)  & 116.84(4) & $-$0.098(025) &  $-$22.89892(5)  \\
    & 0.60 & 1459 &   1605        & 1.20761(14) & 1.09886(04)  & 116.98(5) & $-$0.091(023) &  $-$22.90489(6)  \\
    & 0.40 & 3755 &   2566        & 1.20732(23) & 1.09886(05)  & 116.99(4) & $-$0.057(235) &  $-$22.90761(6) \\
S1  & 0.80 &  782 &   1167        & 1.33888(11) & 1.08562(06)  & 119.40(07)& 34.061(104) &  $-$22.76474(6) & 3.651(2) & 3.652(8) \\
    & 0.60 & 3106 &   2037        & 1.33971(36) & 1.08630(06)  & 119.03(20)& 34.778(051) &  $-$22.77342(5) & 3.577(2) & 3.624(8)\\
    & 0.40 & 9820 &   4031        & 1.33713(31) & 1.08641(06)  & 119.23(11)& 33.522(079) &  $-$22.77682(6) & 3.563(2) & 3.611(8)\\
\hline
\end{tabular}
\end{table*}

We perform four iso-PT2 optimization tests. Starting from a VMC geometry
of Table~\ref{tab:form-geo_S0}, we generate four ground-state wave functions
targeting different values of PT2 correction and further optimize the structure
with these wave functions.  We follow a similar procedure for the excited state
starting from an out-of-plane VMC excited-state geometry of Table S4.  For the
ground state, the resulting bond lengths and angles remain within 1 m{\AA} and
1$^\circ$, respectively, of the CC3 values for all four wave functions.  In
the excited state, the CO bond undergoes a marginal lengthening
and the out-of-pane angle slightly increases by 2-3$^\circ$
becoming closer to the CC3 angle. The adiabatic excitation energies are computed as
differences between iso-PT2 ground and excited-state energies on the corresponding equilibrium
structures, and are highly stable irrespective of the size of the wave function: the VMC values lie
between 3.55-3.56 eV and the DMC ones are somewhat higher and compatible within statistical error with the CC3 estimate. The iso-variance tests yield
very similar results as the iso-PT2 ones and the procedure is therefore also a viable route to compute excited-state structures
and adiabatic energies.

\subsection{Thioformaldehyde}

\begin{table*}[htb]
\caption{
Optimal VMC ground- and excited-state bond lengths (\AA) and bond
angles (deg), and VMC and DMC adiabatic excitation energies (eV)  of thioformaldehyde obtained using Jastrow-CIPSI wave functions with
targeted PT2 correction (iso-PT2). The VMC energy (a.u.) is averaged over the
last 40 iterations of geometry optimization. o.o.p denotes the out of plane
angle of the SO bond.}
\label{tab:thioform-geo}
\begin{tabular}{lcccccccccc}
\hline
State & $\delta$E$_{\rm PT2}$ &  No.\ det.   &   No.\ par.   &   CS       &   CH        &    H-C-H    &   o.o.p    & $\langle$E$_{\rm VMC}\rangle$  & $\Delta$E$_{\rm adia}^{\rm VMC}$ & $\Delta$E$_{\rm adia}^{\rm DMC}$ \\
\hline
S0 &  0.17  & 702   &  995 & 1.62218(14) & 1.08427(04) & 116.869(53) &  $-$0.042(50) & $-$17.05044(4) \\
   &  0.10  & 2500  & 1662 & 1.62250(16) & 1.08469(05) & 116.862(32) &  \;\; 0.070(36) & $-$17.05570(4) \\
   &  0.07  & 5712  & 2810 & 1.62200(11) & 1.08464(06) & 116.786(34) &  $-$0.058(46) & $-$17.05715(6) \\
   &  0.045 & 14218 & 5650 & 1.62207(10) & 1.08480(07) & 116.756(26) &  $-$0.041(53) & $-$17.05817(3) \\
 & \multicolumn{3}{l}{CC3~\cite{budzak2017}}    & 1.619       & 1.083        & 116.1       &   0.000    \\
 & \multicolumn{3}{l}{Expt.~\cite{judge1979,steer1981,clouthier1983}}& 1.611--1.614 & 1.093--1.096      & 116.2--116.9   &  0.000     \\[1.5ex]
S1 &  0.17  &  858  & 764  & 1.72393(27) & 1.07879(07) & 121.309(163) &  $-$0.024(46)     & $-$16.97123(4) & 2.155(2) & 2.152(6) \\
   &  0.10  & 3982  & 1820 & 1.72514(04) & 1.07895(04) & 121.388(27)  &  $-$0.106(79)     & $-$16.97809(5) & 2.112(2) & 2.131(6) \\
   &  0.07  & 8418  & 3110 & 1.72309(07) & 1.07909(05) & 121.149(26)  &   \;\; 0.115(79)  & $-$16.97998(4) & 2.100(2) & 2.119(6) \\
   &  0.045 & 16866 & 5475 & 1.72182(09) & 1.07921(03) & 121.144(28)  &   \;\; 0.041(85)  & $-$16.98090(3) & 2.103(1) & 2.112(6) \\
& \multicolumn{3}{l}{CC3~\cite{budzak2017,loos2018-2}} & 1.709        & 1.078        & 120.2        &  0.0        &               & 2.112 \\
& \multicolumn{3}{l}{Expt.~\cite{judge1979,jensen1982,clouthier1983,dunlop1991}} & 1.682--1.708   & 1.077--1.093  & 116.8--121.6 &  0.0--8.9   &        \\
\hline
\end{tabular}
\end{table*}

Also for thioformaldehyde, we investigate the VMC convergence of the ground-state
optimization with three ground-state wave functions used in the computation of the
vertical excitation energies. As shown in Table \ref{tab:thioform-geo},  we are
able to obtain converged bond lengths and bond angles within less than 3 m{\AA}
and 1$^\circ$ of the CC3 values already with the smallest set of 702 determinants
and a maug-cc-pVDZ basis.  Energetically, we gain much less than 1 mHartee in all
three cases upon optimization starting from the CC3 structure.

In the first singlet excited state, unlike its oxygen counterpart, thioformaldehyde
remains planar and the excited state does not change character.  We therefore start
by simply relaxing the structure with one of the excited-state wave functions
employed in the estimation of the vertical excitation energy. To proceed in the
structural optimization, we then generate new CIPSI expansions following the iso-PT2
scheme and matching the PT2 corrections used in the corresponding ground-state
calculations.  As shown in Table \ref{tab:thioform-geo}, despite testing a large
range of expansions, there is no significant variation in the length of the CS
bond, which is 10 m\AA\ longer than the CC3 value.  The adiabatic excitation energy
is again rather stable across all cases and within 0.01-0.04 eV of the CC3 estimate.


\section{Discussion}
\label{sec:discuss}

We have demonstrated the excellent performance of QMC in the accurate computation
of the vertical excitation energies of two small but theoretically challenging
systems, formaldehyde and thioformaldehyde. Using fully optimized Jastrow-CIPSI
wave functions where the determinantal components are constructed to yield a similar
PT2 correction for both states, we are able to obtain VMC excitation energies
compatible within less than 0.02 eV of the extrapolated FCI values using compact
expansions of a few thousand determinants  and a minimally augmented double-$\zeta$
basis.  Performing DMC calculations on top of the fully optimized VMC wave functions
leads to a uniform gain across both states, thereby not affecting the VMC estimate
of the excitation. If we compute the excitation energy by matching instead wave
functions with similar variances, we obtain a less robust procedure in the case
of formaldehyde, where relatively large expansions of about 7000 determinants are 
required to recover results within 0.05 eV of our best estimates. 

Next, we have investigated the ability of QMC to obtain accurate ground- and
excited-state structures. In the ground state, we easily obtain geometries in
excellent agreement with those produced by other high-level correlated approaches
such as CC3 and full valence CASPT2, using very compact Jastrow-CIPSI wave functions.
The maximum deviation is a meager 3 m{\AA} in the CS bond of thioformaldehyde.  During the relaxation
of the excited-state structure, we regenerate the determinantal component in our
Jastrow-CIPSI wave functions following two different determinant selection schemes:
i) keeping a roughly constant PT2 energy correction during the optimization and ii)
targeting a fixed value of the CI energy variance. Like the vertical excitation
energies, the excited-state structural parameters show relatively low sensitivity
to the size of the wave function and both selection schemes yield similar
structures with no clear distinction in the convergence properties.
The largest
deviation with respect to the CC3 bond lengths is of about 10 m{\AA} when using
the double-$\zeta$ basis. Geometrical relaxation of the excited state additionally allows
us to estimate the adiabatic excitation energies, which we find to be compatible
within less than 0.05 eV of the CC3 values for all wave function sizes.

To summarize, the use of a CIPSI selection scheme targeting similar PT2 corrections
for the states of interest in combination with the full optimization of the
Jastrow-Slater wave function enables us to obtain extremely stable and accurate
estimates of the vertical excitation energies already at the VMC level, namely,
without the need of DMC.  This can be achieved with very compact wave functions,
demonstrating the accuracy 
of the \quotes{expansion} scheme, which is viable also for larger
systems.  Furthermore, our iso-PT2 protocol to regenerate the wave function during relaxation
along a potential energy surface leads to consistent and high-quality structures
for both ground and excited state and to accurate adiabatic excitation energies.
Our robust estimates of the structural parameters and the vertical and adiabatic excitation
energies with moderate Slater expansions open important prospects for the use of
VMC structural optimization with Jastrow-CIPSI wave functions as an efficient
and reliable approach to characterize excited-state potential energy surfaces.

\section*{Acknowledgement}

This work is part of the Industrial Partnership Programme (IPP) ``Computational
sciences for energy research'' of the Netherlands Organisation for Scientific
Research (NWO-I, formerly FOM). This research programme is co-financed by Shell
Global Solutions International B.V. This work was carried out on the Dutch
national supercomputer Cartesius with the support of SURF Cooperative, and
using HPC resources from CALMIP (Toulouse) under allocation 2018-0510.  J.F.
acknowledges the Deutsche Forschungsgemeinschaft (DFG) for financial support
(Grant FE 1898/1-1). The authors thank Dr. Pierre-Francois Loos for useful
discussion and data on the coupled cluster geometries. The authors declare no
competing financial interest.

\section*{Suppinfo}

CIPSI energies and FCI/maug-cc-pVDZ extrapolation for formaldehyde and thioformaldehyde;
performance of \quotes{truncation} scheme in VMC; fits of ground- and excited-state VMC energies 
as a function of the total number of determinants and as a function of the energy variance;
DMC time-step extrapolation; intermediate excited-state geometry optimizations.

\bibliography{paper_optgeo_CIPSIQMC}

\begin{mcitethebibliography}{78}
\bibitem[Jas()]{Jastrow}
As Jastrow factor, we use the exponential of the sum of two fifth-order
  polynomials of the electron-nuclear and the electron-electron distances,
  respectively, and rescale the inter-particle distances as $R=(1-\exp(-\kappa
  r))/\kappa$ with $\kappa$ set to 0.6 a.u. We employ different
  electron-nucleus Jastrow factors to describe the correlation of an elecron
  with C and H. The total number of free parameters to be optimized in the
  Jastrow factor is 17 for the systems considered here.\relax
\mciteBstWouldAddEndPunctfalse
\mciteSetBstMidEndSepPunct{\mcitedefaultmidpunct}
{}{\mcitedefaultseppunct}\relax
\EndOfBibitem
\bibitem[Det()]{DetTot}
As the determinants are selected altogether, the total number of determinants
  represents the size of the determinantal space in which the Hamiltonian is
  diagonalized to obtain both states. Note that as the states are in different
  symmetries, the number of non-zero determinant (or CSF) coefficients for each
  state is different.\relax
\mciteBstWouldAddEndPunctfalse
\mciteSetBstMidEndSepPunct{\mcitedefaultmidpunct}
{}{\mcitedefaultseppunct}\relax
\EndOfBibitem
\end{mcitethebibliography}

\end{document}